\documentclass[preprint, 
               notlongauthorlist 
              ]{nsr}

\usepackage{soul}        
\usepackage{lineno, blindtext} 
\usepackage{amsfonts}

\usepackage{placeins}
\usepackage{subcaption}
\makeatletter
\providecommand{\@runauth}{}
\makeatother
\volume{00}

\artnum{00}

\firstpage{1}

\datesubmitted{ 18 June 2026}
\datereceived{18 June 2026} 
\daterevised{5 July 2026}
\dateaccepted{3 August Year} 
\datepublished{13 August Year}

\doinum{doi/number}

\copyrightyear{2026}

\author[1]{I. V. Yeletskikh\footnote{Corresponding author e-mail address: \tt ivaneleckih@jinr.ru}$^{\orcidlink{0000-0003-0586-7052}}$}
\author[1]{A. O. Vasyukov\footnote{Corresponding author e-mail address: \tt avasyukov@jinr.ru}$^{\orcidlink{0000-0003-2460-1276}}$}

\affil[1]{Joint Institute for Nuclear Research, Dubna, 141980, Russia}

\title{Multivariate amplitude analysis of the cascade particle decays based on the Nearest Neighbors fitting}

\begin{document}


\maketitle

\begin{abstract}
\noindent
    The Nearest Neighbors estimation of likelihood is implemented for the multivariate amplitude analysis of the cascades of particle decays. In this approach Monte Carlo simulated events that are used to describe data are assigned weights dependent on the parameters of the theoretical fitting model. Each of such weighted events is considered as contributing to N-dimensional probability density function in it's neighborhood in the parameter space. No analytic p.d.f. or cpu-intensive re-generation of N-dimensional p.d.f. at each fit step are required. 
    The method allows accurate accounting for reconstruction effects, demands modest cpu resources and can be effectively multi-threaded. General setup of the fit as well as toy example of the  
    amplitude analysis of B-meson decays are demonstrated.
\end{abstract}

\begin{keyword}
nearest neighbor, multivariate analysis, multidimensional fit, cascade decay
\end{keyword}


\FloatBarrier
\section{Introduction}\label{sec1}

Advanced approaches to scientific data processing are often based on Multi-Variate Analysis (MVA), i.e., analysis of many data characteristics simultaneously, which allows better accuracy of measurements, observation of complex effects, correct accounting for correlations, etc.
With a number of popular MVA methods, like Principal Component Analysis or Machine Learning applications, the reduction of the dimensionailty of the problem is achieved, however, it relies on the \textit{a priori} known properties of data -- either from analytic description or from simulation in many dimensions.
However, when model components, e.g., signals and backgrounds, are tangled in a complex manner and cannot be factorized -- model properties have to be defined from fits. In this case the full-fledged N-dimensional fits of model to data have to be applied.

\begin{figure}[h]
\centering
\includegraphics[width=0.9\textwidth]{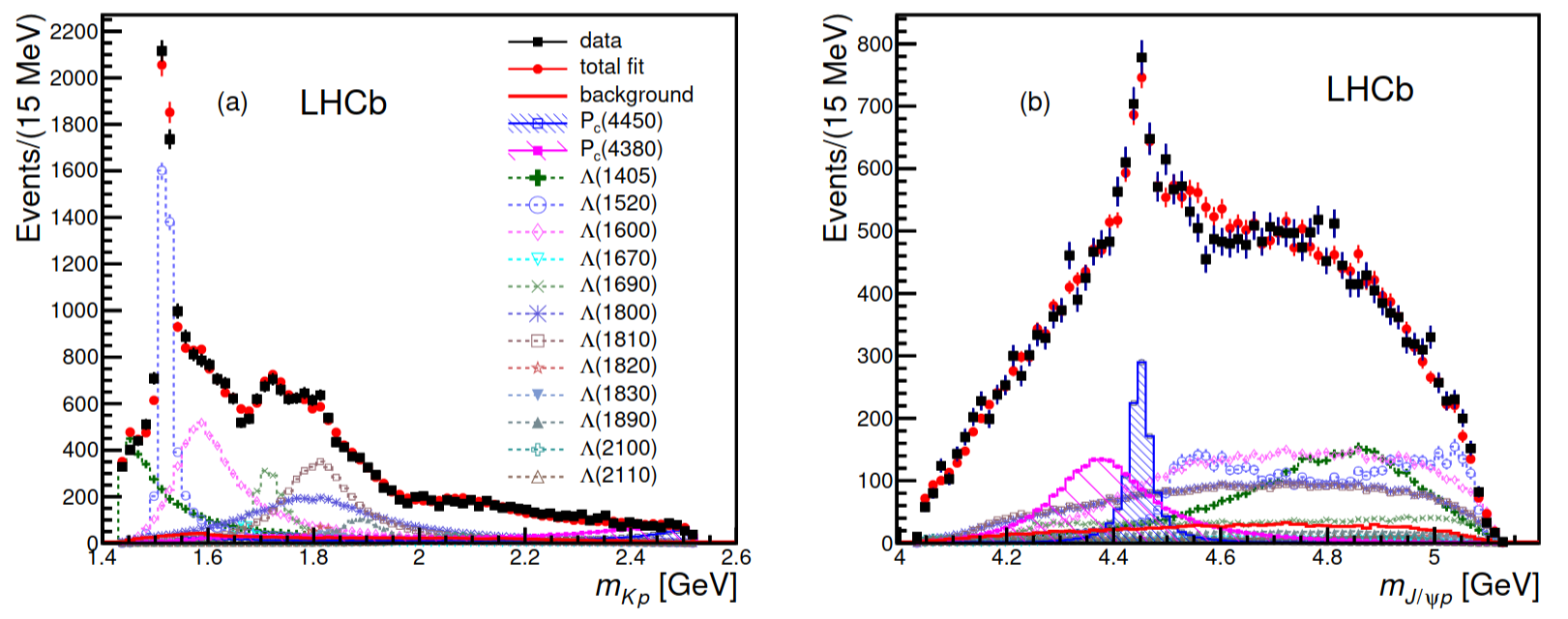}
\caption{Results of multidimensional unbinned likelihood fit of LHCb data for selected $\Lambda_b\rightarrow J/\psi p K^-$ decays \cite{PhysRevLett.115.072001}. }\label{fig1}
\end{figure}

\begin{figure}[h]
\centering
\includegraphics[width=0.6\textwidth]{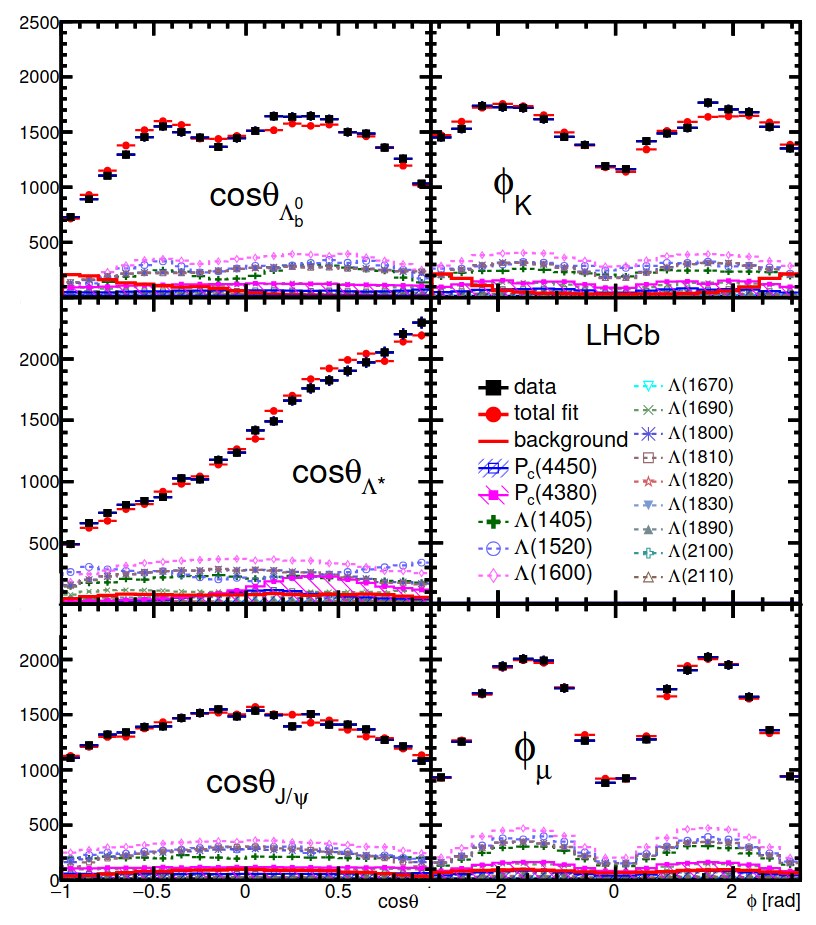}
\caption{Results of multidimensional unbinned likelihood fit of LHCb data for selected $\Lambda_b\rightarrow J/\psi p K^-$ decays \cite{PhysRevLett.115.072001}.  }\label{fig2}
\end{figure}

When an analytical representation of the model expressed via probability distribution function (p.d.f.) is available -- the N-dimensional unbinned likelihood fits are possible with no limitations on the number of dimensions. However, not only precise analytic models are rarely available for complex data, some additional assumptions or approximations are usually needed. In particular, the uncertainties in the raw data, due to reconstruction effects in the detectors, are either ignored or approximated by folding p.d.f. with distributions describing such errors, often a Gaussian distribution. 
A good example of applying analytic model in multidimensional fits is the LHCb collaboration (CERN) study \cite{PhysRevLett.115.072001} of pentaquark hadrons in $\Lambda_b$-baryon decays. The analytic p.d.f. is formed in this work as:

\begin{equation}
\mathcal{P} = \frac{1}{I(\omega)} | M(\omega)|^2 \Phi(\omega) \epsilon(\omega).   \label{eq1}
\end{equation}
where $\omega$ is a set of fit parameters (pole masses and widths of resonances, coupling constants), $I(\omega)$ -- normalization integral, $\Phi(\omega)$ -- phase space distribution (function of the 4-momenta of decay products), $\epsilon(\omega)$ -- efficiency/acceptance factor, $M(\omega)$ -- sum of amplitudes for all signals/backgrounds included in the decay model. 
The p.d.f. of this form is used to perform 6-dimensional fits of the one invariant mass and 5 angular variables. $\omega$ parameters are initially unknown, so fit alternates them to adjust resulting p.d.f. to data points in 6D kinematic space.  
Figures \ref{fig1}, \ref{fig2} demonstrate results of such fits.

Contributions from processes other than $\Lambda_b\rightarrow J/\psi p K^-$ decays, e.g., combinatorial background, are difficult to be described by analytic functions in this scheme. In the LHCb analysis, some of these backgrounds are subtracted from fitted distributions using data-driven techniques, other are neglected or included in systematic uncertainties. 
Many experimental problems suffer from high backgrounds that can be accurately modeled by Monte Carlo (MC) methods only. It makes practically impossible the usage of analytic p.d.f. in N-dimensional (ND) fits. 

When analytic p.d.f. is unknown or inaccurate due to above reasons, one may generate ND p.d.f. from discrete MC simulated events. For example, in the widely used 'RooFit' \cite{roofit0} package there are several methods (like \textit{RooNDKeysPdf}), which process the 'ntuple' (a set of simulated variables in event-by-event format) as input and provides kernel density estimation by approximating event distributions with a superposition of N-dimensional Gaussian functions.
An example of data analysis that uses this technique is BESIII Collaboration study \cite{PhysRevD.111.052012} of $\psi(3686)$-meson decays to $\Sigma^0 \bar{\Sigma}^0\phi$ final states. MC simulated events of the latter decays are converted to 3-dimensional p.d.f. via \textit{RooNDKeysPdf} to perform fit of the $\phi\rightarrow K^+K^-$ and $\Sigma^0\rightarrow \gamma\Lambda$, $\bar{\Sigma^0}\rightarrow \gamma\bar{\Lambda}$ decays.

\begin{figure}[h]
\centering
\includegraphics[width=0.98\textwidth]{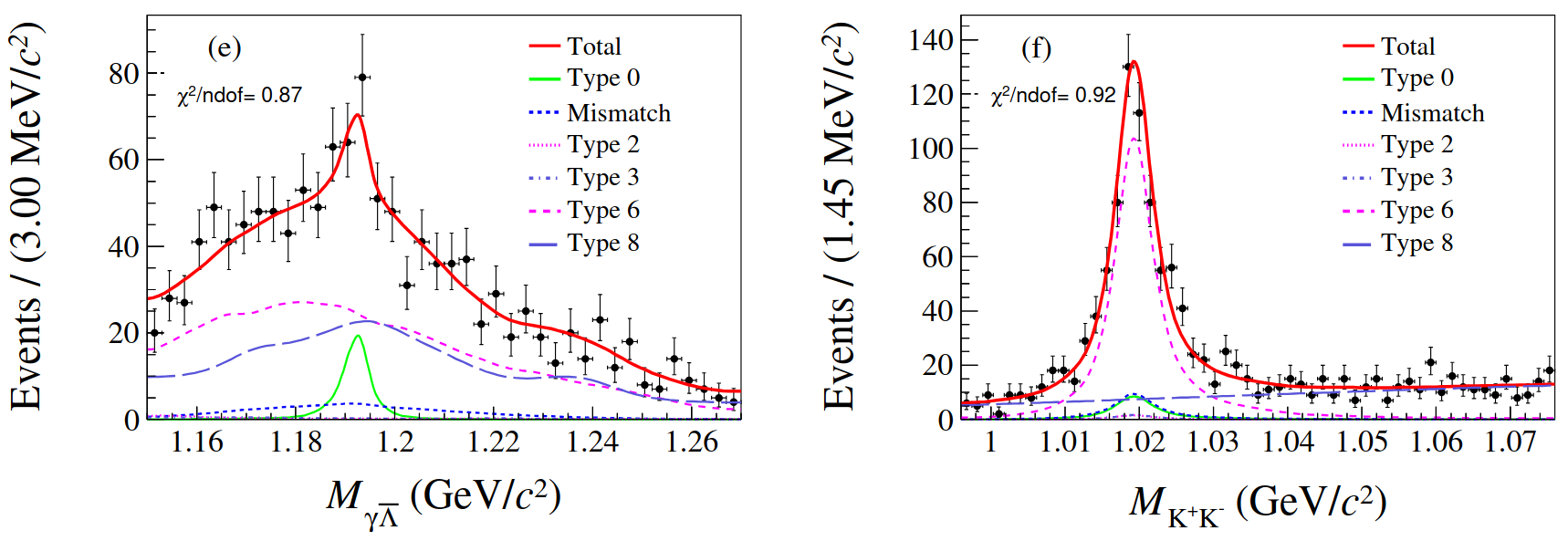}
\caption{Results of 3-dimensional unbinned likelihood fit of BESIII data of  $\psi(3686) \rightarrow \Sigma^0 \bar{\Sigma}^0\phi$  decays \cite{PhysRevD.111.052012}. $\bar{\Sigma^0}\rightarrow \gamma\bar{\Lambda}$ selected candidates (left)  and $\phi\rightarrow K^+K^-$ selected candidates (right) are shown.}\label{fig3}
\end{figure}

In the latter analysis, positions and shapes of signals (masses, widths) are known and fixed -- only yields are fitted from data. In the opposite case (when masses/widths of signals are unknown or signals/backgrounds are tangled via complex interference effects), generation of the new p.d.f. would be needed at each fit step -- after any parameter affecting signal/background shapes has been changed. Such 'regeneration' may take a lot of computing resources, especially so with high number of fit dimensions.

Another possible option for implementing N-dimensional fits is the $\chi^2$ fit of model to data both represented via histograms. Number of dimensions in this case is usually limited by 2-3. Higher dimensionality would typically require developing custom histogram classes and methods to optimize memory and cpu consumption. Also, to assure sufficient number of entries in each bin (at least $>$ 10) the non-uniform histogram binning (or merging neighboring bins in more general case) is used. Definition of such custom binning may become a separate optimization problem. Rough binning lowers sensitivity to data properties, fine binning increases data and model uncertainties in each bin.
The ND $\chi^2$ approach for data description quality estimation has been used in BELLE collaboration analysis \cite{PhysRevD.90.112009} of $B^0$-meson to $J/\psi K\pi$ decays. In this study, the 2D $\chi^2$ is not used during fits -- instead it is applied to assess goodness of 2D Dalitz plots that were binned non-uniformly to contain at least 25 entries in each bin.

\section{Nearest Neighbors likelihood fit}\label{sec2}

With the aim to overcome limitations of the above mentioned approaches, the goodness-of-fit estimation using 'distance to nearest neighbor' may be considered. It was demonstrated \cite{Schilling01091986, HenzeNorbert} that the 'nearest' (or the 'closest') neighbor method can be used for the multivariate comparison of two samples of events.
In this method, the value of the p.d.f. for each data event is estimated based on one (or several) kinematically closest MC events -- without explicit (analytic or numeric) p.d.f. definition prior to fit and at each fit step. 

In one of the various applications of this method it is assumed that each MC event makes certain individual contribution to the total p.d.f. (at a given data event position) depending on the distance between data and MC positions. This contribution may be estimated by N-dimensional Gaussian function with the center at the MC event position and N widths that are chosen to provide smoothness of the resulting total p.d.f.
Thus, the value of resulting p.d.f. at the $i$-th data point, having $\vec{\theta}_i$ coordinates in ND space, is defined as a plain sum:

\begin{equation}
\mathcal{P}_i = \frac{1}{ \sum_{k} w_k } \sum_{j} w_j \times Gauss_{ND}(\vec{\theta}_{i} - \vec{\theta}_{j}, \vec{\sigma}),   \label{eq2}
\end{equation}
where $\vec{\theta_j}$ are coordinated of MC events, $w_j$ -- weights of MC events, $Gauss_{ND}$ is the multivariate Gaussian distribution. The normalization of p.d.f. is accomplished by normalization factor $\sum_{k} w_k$, which is a sum of MC event weights.
$w_j$ are typically composed of some theoretical model weights multiplied by detector scale/efficiency factors, normalization factors, etc. $w_j$ are computed using generator-level (also called 'truth-level') parameters of the given MC event, while the $\vec{\theta}_{i} - \vec{\theta}_{j}$ distances are taken for the reconstructed values (also called 'detector level'). It allows the full-fledged estimation of the detector effects in p.d.f.
In the simplest case, $Gauss_{ND} = \prod_{dim} Gauss(\theta_{i,dim} - \theta_{j,dim}, \sigma_{dim})$ with $dim$ enumerating dimensions of the parameter space, and $Gauss$ to be plain Gaussian normalized to unity integral. $\vec{\sigma}$ represents covariance vector for the fitted variables. $\vec\sigma$ may be chosen to reflect characteristic scales of reconstruction uncertainties for each dimension in the fit. In practice $\vec\sigma$ may be increased in those parameter space regions with lower MC event statistics bearing in mind that the typical size of structures in data (like widths of signals) in each dimension must not be much smaller than $\vec\sigma$.

Note that there is no need to calculate the sum in equation (\ref{eq2}) over all MC events. With good practical accuracy one may take only those MC events that are located within 4---5$\sigma$ proximity to the $i$-th data event for further cpu optimization.
Moreover, in almost all practical cases, when kinematics of MC events is not varied during fit, all values of $Gauss_{ND}$ functions stay the same for given data and MC samples. They can be pre-computed and stored in memory with only $w_j$ MC event weights being varied during fit. Recalculation of $w_j$ and resummation of individual p.d.f. contributions with different weights $w_j$ is done at each fit step. It substantially optimizes cpu consumption. Moreover, calculation of $w_j$ is very well parallelized to run on multiple cpu or gpu cores.

The final ND likelihood function is obtained using the p.d.f. calculated at each data point using formula (\ref{eq2}):

\begin{equation}
log\mathcal{L} = \sum_{i} log(\mathcal{P}_i) \times v_i,  \label{eq3}
\end{equation}
where $v_i$ are data event weights if they are different from 1.


\section{Fit setup and tests}\label{sec3}
Let us consider few practical examples to demonstrate performance of nearest neighbor likelihood in multiple dimensions.
\subsection{Fit of simple ND functional form}\label{subsec31}
As a very basic test of the suggested approach, the 5-dimensional pseudodata is generated with simple analytic p.d.f.:
\begin{equation}
\mathcal{P}_5 = p_0 + p_1x_1^2 +  p_2x_2^2 + p_3x_3^2 + p_4x_4^2 + p_5x_4x_5,   \label{eq4}
\end{equation}
where $x_i$ -- variables with randomly generated values in $[0,2]$ range, $p_i$ -- fit parameters, that have to be found from fit. For data generation, the values $x_i=[0.5, 0.35, 0.1, 0.03, 0.4, 0.2]$ were chosen. Correlations in data distribution are 0 in dimensions 1, 2, 3, while dimensions 4 and 5 are correlated to test if fit copes with correlated variables.
To describe data, six sets of MC events is generated, using the following p.d.f.'s: $1.0$, $x_1^2$,  $x_2^2$, $x_3^2$, $x_4^2$, $x_4*x_5$. 118K of pseudodata events were generated, while each set of MC events contains 50-100K of events.

\begin{figure}[h]
\centering
\includegraphics[width=0.49\textwidth]{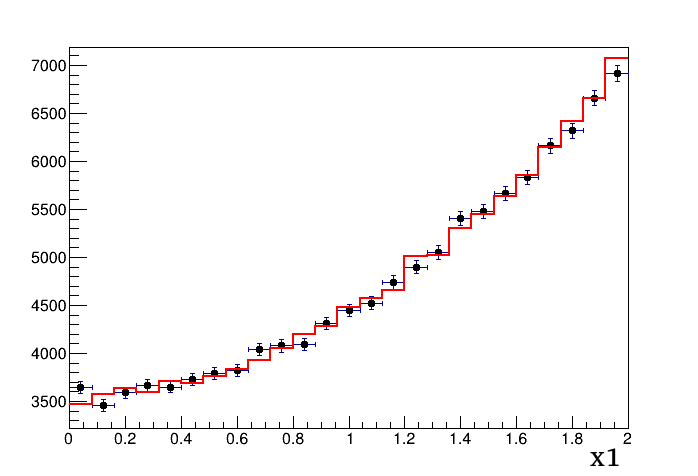}
\includegraphics[width=0.49\textwidth]{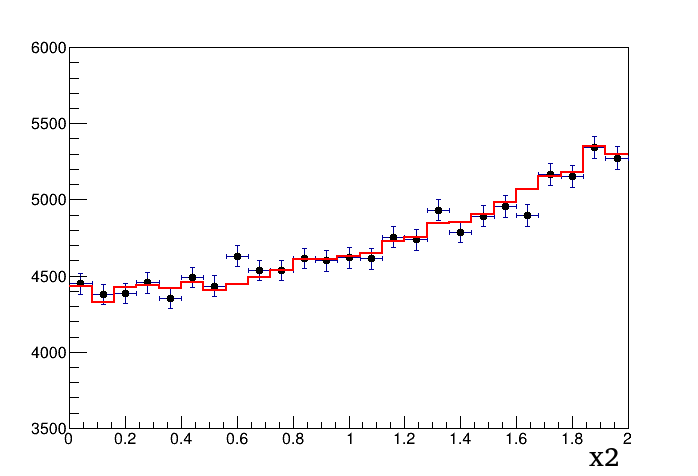}
\includegraphics[width=0.49\textwidth]{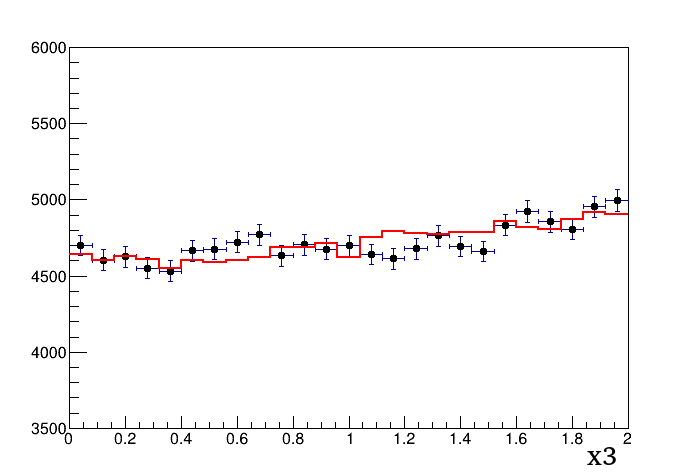}
\includegraphics[width=0.49\textwidth]{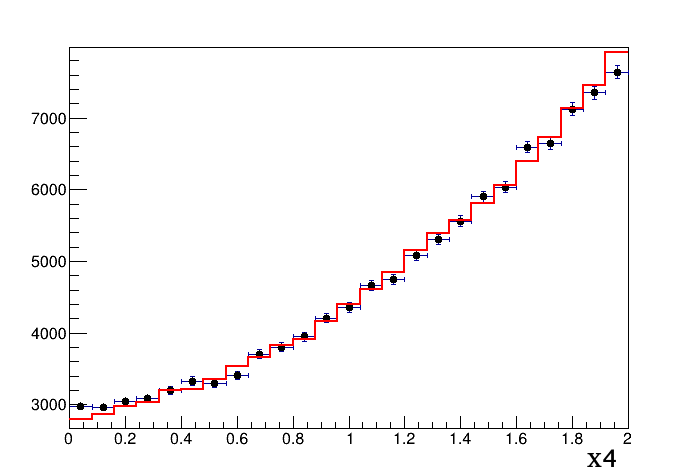}
\includegraphics[width=0.49\textwidth]{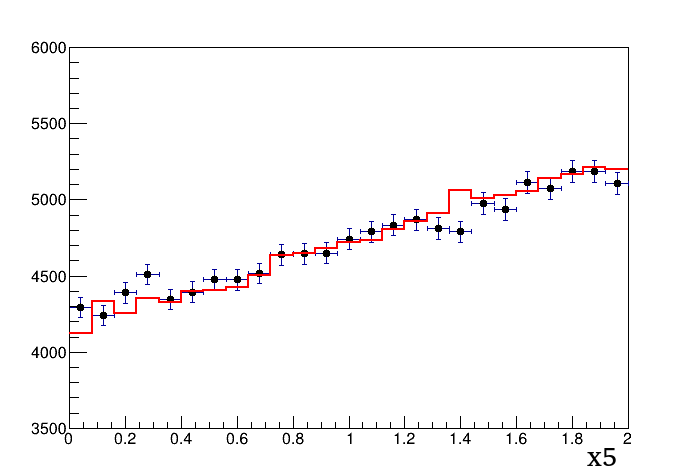}
\caption{Results of 5-dimensional unbinned likelihood fit of pseudo data. X-axis at each plot corresponds to value of each variable, while Y-axis demonstrate numbers of events -- projections  to 5 axes are shown (black dots) together with the fitted model (red line).}\label{fig4}
\end{figure}

Parameters found by the fit are the following: $[0.52\pm0.11, 0.44\pm0.09, 0.10\pm0.02, 0.04\pm0.01, 0.31\pm0.07, 0.19\pm0.04]$. All parameters are found with adequate accuracy.
Figure \ref{fig4} shows projections of data distribution to 5 axes together with the fitted model.

These results demonstrate that our version of the nearest neighbor method is capable for ND fits without the need for analytic p.d.f. or cpu-consuming generation of p.d.f. at each fit step. No tuning of the implicit parameters (like non-uniform binning configuration in $\chi^2$ fits) is needed as well. 
Trivial requirements for such fits to guarantee reasonable likelihood statistics is providing sufficient event statistics of MC simulation. A set of implicit parameters -- Gaussian widths ($\vec{\sigma}$) are chosen to ensure that reasonable number of MC events (from tens to several thousands in a 3-sigma proximity) contribute to p.d.f. at each data point. It saves fit from falling into local minima related to model fluctuations. 

\subsection{Example fit of $B^0\rightarrow J/\psi K^+\pi^-$, $\bar{B}^0\rightarrow J/\psi K^-\pi^+$ decay model}\label{subsec32}

\begin{figure}[h]
\centering
\includegraphics[width=0.7\textwidth]{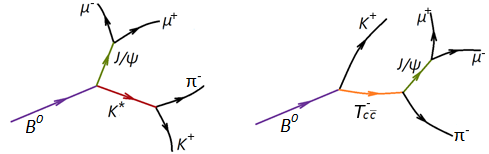}
\caption{Diagrams of $B^0\rightarrow J/\psi K^+\pi^-$ cascade decays. Decay chain via $K^*$ on the left, decay chain via tetraquarks $T^{\pm}_{c\bar{c}}$ on the right.}\label{fig50}
\end{figure}

As a more realistic implementation of ND closest neighbor fit, let's consider a physical model of $B^0$-meson decays -- $B^0\rightarrow J/\psi K^+\pi^-$ and $\bar{B}^0\rightarrow J/\psi K^-\pi^+$. In this case, the initial state ($B^0$ or $\bar{B}^0$) decays into final state of three particles -- $J/\psi$-meson, kaon ($K^{\pm}$) and pion ($\pi^{\mp}$). The complexity of these decays is due to possible mediation by different resonant states -- $K^*$ or $T^{\pm}_{c\bar{c}}$. Decays proceed as $B^0\rightarrow J/\psi K^*$ with $K^*\rightarrow K^+\pi^-$ afterwards, $B^0\rightarrow T^{-}_{c\bar{c}}K^+$ with $T^{-}_{c\bar{c}}\rightarrow J/\psi\pi^-$ afterwards, see Fig. \ref{fig50}. It should be stressed also that a number of different $K^*$ states and different $T^{\pm}_{c\bar{c}}$ states with different pole masses, widths and couplings may contribute to such $B^0$ decays.

Each type of decay chain may be described by theoretical amplitude -- a complex-number function depending on the decay parameters (momenta of particles in the final state and angles between them). To compute the total matrix element of $B^0\rightarrow J/\psi K^+\pi^-$ all individual resonant amplitudes scaled by complex decay constants are summed coherently. 

In practice, exact values of decay constants for these different types of decay chains are hard to estimate from theory. Moreover, parameters other than decay constants (e.g., masses and widths of some states) are unknown $a priori$. 
Assume also that our particle detector can not distinguish light hadron flavors (in this case - kaons and pions), so $B^0\rightarrow J/\psi K^+\pi^-$ and $\bar{B}^0\rightarrow J/\psi K^-\pi^+$ are mixed together. This is often the case in real experimental studies.

Thus, the total matrix element is a non-trivial mix-up of different components. It means that the total p.d.f. can not be factorized with respect to $T^{\pm}_{c\bar{c}}$ signals and $K^*$ backgrounds and has to be redefined at each fit step depending on parameter values.


For this example, Pythia 8.1 MC generator \cite{SJOSTRAND2008852} is used to simulate $B^0$-, $\bar{B}^0$-meson production in pp-collisions with an energy of 13 TeV in the center of mass system and their decays to $J/\psi$, kaon and pion according to the phase-space distribution. These 'phase-space' events are weighted by analytic helicity amplitudes modeling different intermediate resonances. Parameters of helicity amplitudes are fixed to some initial values in pseudodata generation. Reconstruction of generated events is accomplished using the simulated model of a typical modern experimental facility.

137k pseudodata events have been generated (68.5K $B^0\rightarrow J/\psi K^+\pi^-$ and 68.5K  $\bar{B}^0\rightarrow J/\psi K^-\pi^+$) with two intermediate $K^*$ resonances $K^*$(1600) (spin 2) and $K^*$(1950) (spin 0) and one $T_{c\bar{c}}$(4200) state with spin 2 contributing to decays. The masses, widths and spin-parity properties are given by the values listed in Table \ref{tab:paramZcsummary}. 

Sample of events that is used to fit pseudodata is based on the same 'phase space' Pythia 8.1 generation. During fit, parameters of the helicity amplitudes may vary, affecting the weight assigned to each 'phase space' event. The model used to fit pseudodata has 7 free parameters of helicity amplitudes and includes 950K of $B^0\rightarrow J/\psi K^+\pi^-$ plus 950K of $\bar{B}^0\rightarrow J/\psi K^-\pi^+$ events. Masses and widths for $K^*$ states are fixed in the model, while mass and width of $T_{c\bar{c}}$ state are free. Another 5 free parameters of the fit correspond to three complex decay constants -- one for each intermediate state minus one parameter of the common complex phase being fixed.

Fit is run by 'MIGRAD' algorithm using likelihood estimated according to \ref{eq3} of the 6-dimensional distribution of kinematic variables (invariant masses): $m(J/\psi\pi^+)$, $m(J/\psi\pi^-)$, $m(J/\psi K^+)$, $m(J/\psi K^-)$, $m(K^+\pi^-)$, $m(K^-\pi^+)$. 
Both 'kaon' and 'pion' mass assignments are tested for each hadron track to reflect the typical situation with absence of hadron flavor identification at the experiment.
All $\sigma$ values are set to 10 MeV, which is typical modern detector accuracy for reconstruction of these types of particles.
The fitting code takes ~10s per iteration running in multi-threaded mode on a modern 8-core cpu.

\begin{figure}[h]
\centering
\includegraphics[width=0.49\textwidth]{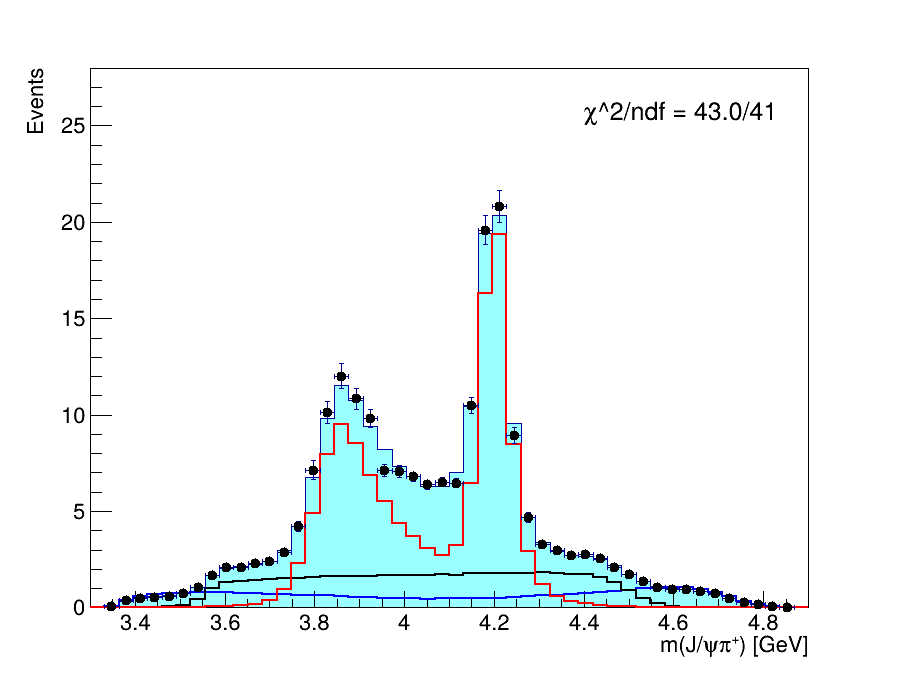}
\includegraphics[width=0.49\textwidth]{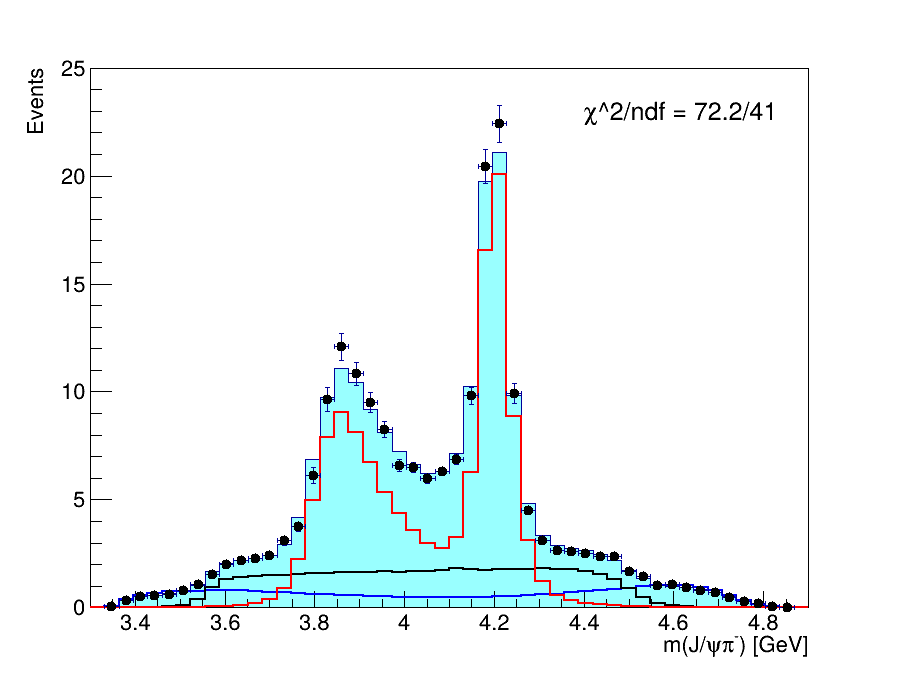}
\includegraphics[width=0.49\textwidth]{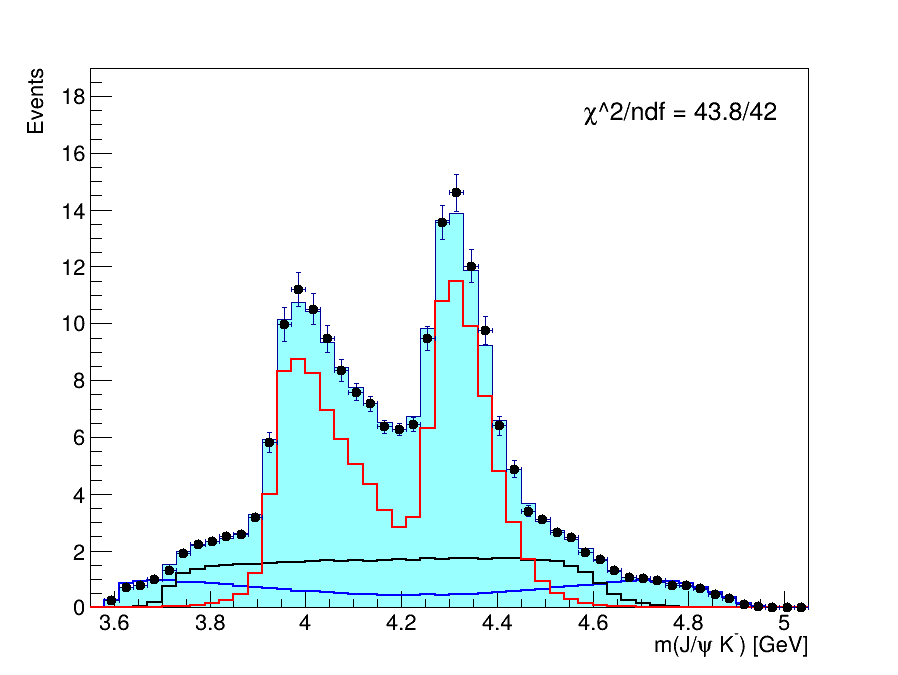}
\includegraphics[width=0.49\textwidth]{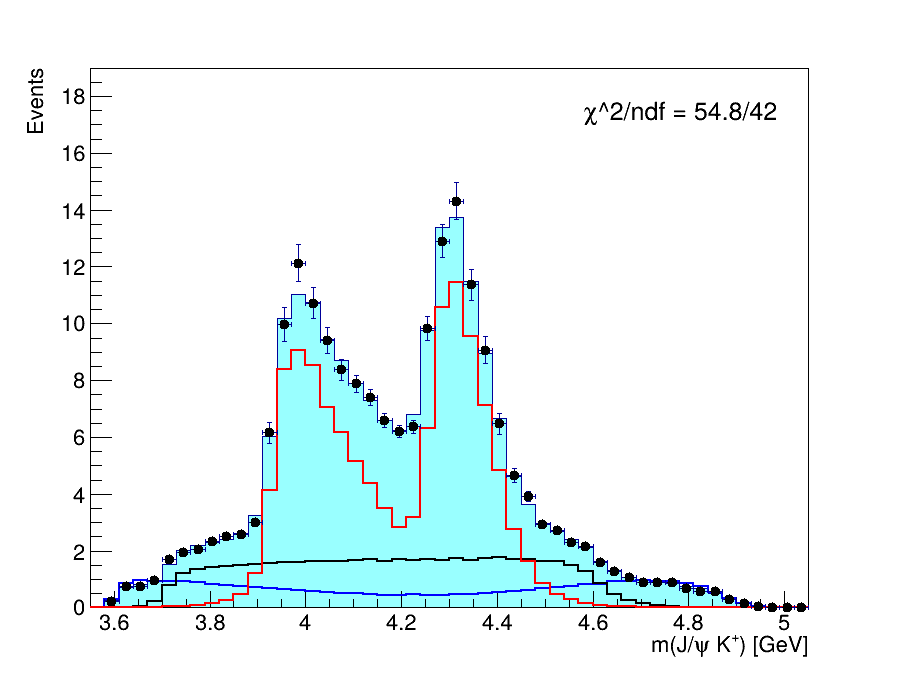}
\includegraphics[width=0.49\textwidth]{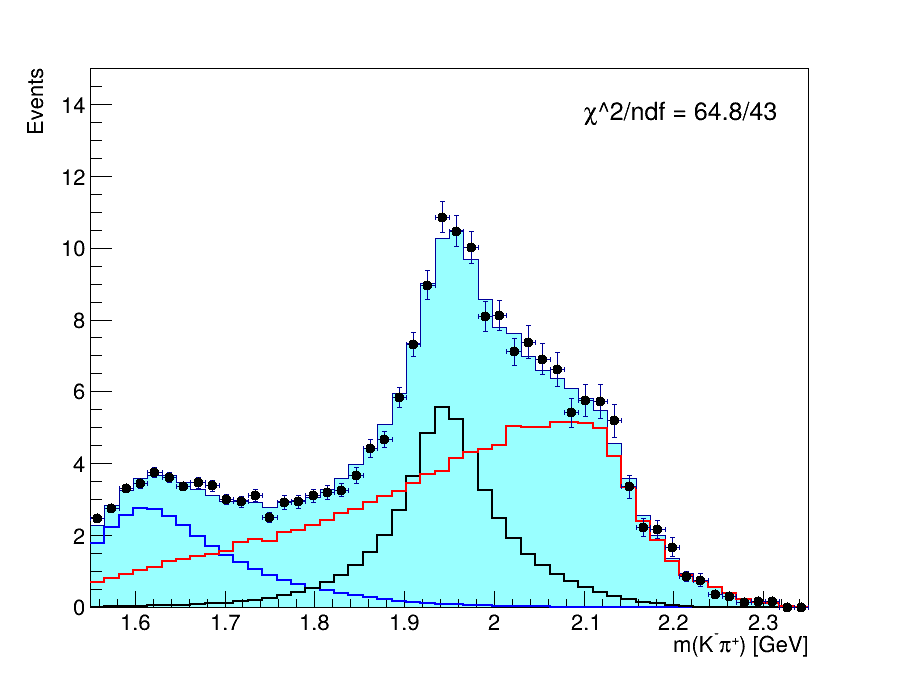}
\includegraphics[width=0.49\textwidth]{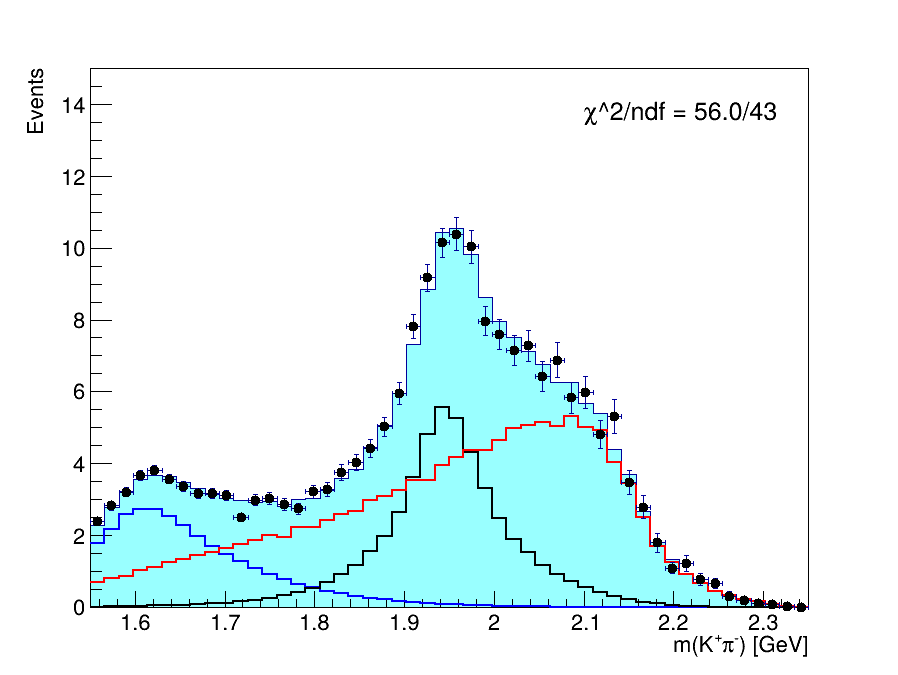}
\caption{Results of 6-dimensional unbinned likelihood fit of pseudodata simulating mixture of $B^0\rightarrow J/\psi K^+\pi^-$ and $\bar{B}^0\rightarrow J/\psi K^-\pi^+$ decays. Projections of data to 6 axes are shown (black dots) together with the fitted model (azure histogram). Contributions of three different amplitudes taken before coherent summation are shown by blue line ($K^*$(1430)), black line ($K^*$(1950)), red line ($T_{c\bar{c}}$(4200)). }\label{fig5}
\end{figure}

Fig. \ref{fig5} demonstrates distributions of pseudodata over all fitted dimensions together with fitted model and contributions of different amplitudes before coherent summation with each other.
Different intermediate states can be seen as peaking structures in the different spectra of these invariant masses.

\begin{table}[h!]
\caption{
Initial values used to generate pseudodata and corresponding values obtained from 6-dimensional fitting.
}
\renewcommand{\arraystretch}{1.4} 
\centering
\begin{tabular}{c|c|c} \hline
Parameter & Initial value & Fitted value \\
\hline
$T_{c\bar{c}}$ mass  & 4200 MeV & 4201 MeV \\ 
$T_{c\bar{c}}$ width & 30.0 MeV & 32.4 MeV \\ 
$T_{c\bar{c}}$ ampl & 20.0 & 20.0 \\ 
$T_{c\bar{c}}$ phase & $\pi$ & 1.01$\pi$ \\ 
\hline
$K^*(1600)$ mass & 1600 MeV & fixed \\ 
$K^*(1600)$ width & 109.0 MeV & fixed \\ 
$K^*(1600)$ ampl & 2.00 & 1.94 \\ 
$K^*(1600)$ phase & 0.0 & fixed \\ 
\hline
$K^*(1950)$ mass & 1944 MeV & fixed \\ 
$K^*(1950)$ width & 40.0 & fixed \\ 
$K^*(1950)$ ampl & 0.030 & 0.029 \\ 
$K^*(1950)$ phase & $\pi/2$ & 0.503$\pi$ \\
\hline
\hline

\end{tabular}
\label{tab:paramZcsummary}
\end{table}

Table \ref{tab:paramZcsummary} summarized reconstructed parameters by the fit. It can be seen that fit outputs model parameters close to that used in pseudodata generation -- not only mass and width of $T_{c\bar{c}}$ state, but also amplitudes and phases of decay constants that reproduce correct picture of interference between different resonances.

\section{Conclusions}\label{conclusion}
The application of the nearest neighbors likelihood fits is tested on the realistic example of amplitude analysis of particle cascades.
Consumption of cpu is modest compared to alternative ND fit approaches. The scheme does not require explicit computation of N-dimentional p.d.f. at each fit step. In comparison to analytic p.d.f. fits, the new method utilizes all powers of MC methods and allows accurate accounting for the effects of particle reconstruction in detector. Number of dimensions is limited by nothing except statistics of simulated events to keep statistical uncertainties of the model small. Test fits show good capability of reproducing parameters of $B^0\rightarrow J/\psi K^{\pm}\pi^{\mp}$ decays in pseudodata.

\FloatBarrier
\bibliographystyle{nsr}


\end{document}